\documentclass[sigconf,natbib=true,review=False, anonymous=False,nonacm]{acmart}

\AtBeginDocument{%
  }

\setcopyright{acmlicensed}
\copyrightyear{2018}
\acmYear{2018}
\acmDOI{XXXXXXX.XXXXXXX}

\newcommand{\nanpy}{NApy }

\newcommand{\data}{\mathbf{X}}

\usepackage{csquotes}
\usepackage{tabularx}
\usepackage{cleveref}
\usepackage{multirow}
\usepackage{rotating}
\usepackage{makecell}   
\usepackage{graphicx}   

\begin{document}

\title{NApy: Efficient Statistics in Python for Large-Scale Heterogeneous Data with Enhanced Support for Missing Data}

\author{Fabian Woller}
\authornote{Both authors contributed equally to this research.}
\email{fabian.woller@fau.de}
\orcid{0000-0001-5492-6819}
\affiliation{%
  \institution{Friedrich-Alexander-Universit{\"a}t Erlangen-N{\"u}rnberg}
  \city{Erlangen}
  \country{Germany}
}

\author{Lis Arend}
\authornotemark[1]
\email{lis.arend@tum.de}
\orcid{0000-0001-7990-8385}
\affiliation{%
  \institution{Technical University of Munich}
  \city{Munich}
  \country{Germany}
}

\author{Christian Fuchsberger}
\email{christian.fuchsberger@eurac.edu}
\affiliation{
\institution{Eurac Research}
\city{Bolzano}
\country{Italy}
}

\author{Markus List}
\email{markus.list@tum.de}
\affiliation{%
  \institution{Technical University of Munich}
  \city{Munich}
  \country{Germany}
}

\author{David B. Blumenthal}
\email{david.b.blumenthal@fau.de}
\affiliation{%
  \institution{Friedrich-Alexander-Universit{\"a}t Erlangen-N{\"u}rnberg}
  \city{Erlangen}
  \country{Germany}
}


\begin{abstract}
Existing Python libraries and tools lack the ability to efficiently compute statistical test results for large datasets in the presence of missing values. This presents an issue as soon as constraints on runtime and memory availability become essential considerations for a particular usecase. Relevant research areas where such limitations arise include interactive tools and databases for exploratory analysis of biomedical data. To address this problem, we present the Python package NApy, which relies on a Numba and C++ backend with OpenMP parallelization to enable scalable statistical testing for mixed-type datasets in the presence of missing values. Both with respect to runtime and memory consumption, NApy outperforms competitor tools and baseline implementations with na\"ive Python-based parallelization by orders of magnitude, thereby enabling on-the-fly analyses in interactive applications. NApy is publicly available at \url{https://github.com/DyHealthNet/NApy}.
\end{abstract}



\begin{teaserfigure}
  \includegraphics[width=\textwidth]{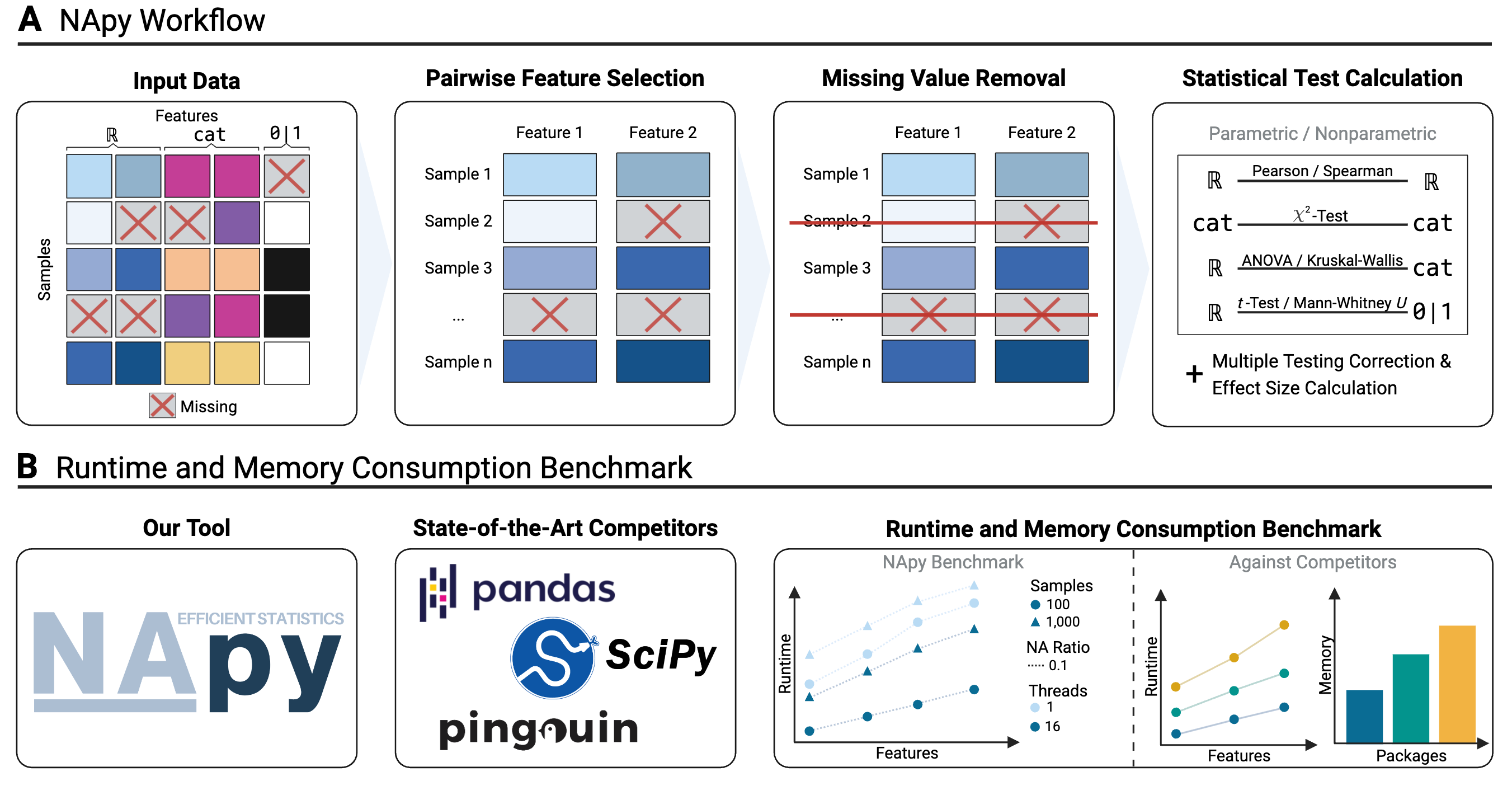}
  \caption{Overview of NApy's workflow (A) and benchmark analyses of runtime and memory consumption (B). NApy enables efficient computation of standard statistical tests in parallel with pairwise deletion of missing values. Benchmarking on both simulated and a population-based cohort study demonstrates that NApy outperforms state-of-the-art competitors by orders of magnitude.}
  \label{fig:teaser}
\end{teaserfigure}

\maketitle

\section{Introduction}\label{sec:introduction}

Statistical analyses in biomedical research areas often deal with datasets consisting of several thousands of samples and variables. Especially population-based cohort studies such as the Cooperative Health Research in South Tyrol (CHRIS) study \cite{Pattaro2015-ok} or the Study of Health in Pomeria (SHIP) \cite{Volzke2011-mm} typically include several thousand study participants. These studies typically encompass diverse health and lifestyle records represented by quantitative (e.g. blood pressure, height), dichotomous (i.e binary, such as sex, presence of a disease), and categorical variables (e.g. body mass index categories, job). Additionally, due to factors like experimental design or the non-applicability of certain measurements to specific subgroups of participants, the population cohort data frequently exhibits a significant amount of missing data \cite{Little2012-vm}.

While, classically, biomedical data analyses on population cohort data  mostly investigate statistical associations between \emph{specific} variables (e.g., correlations between measurements as body mass index and blood pressure in population cohorts), the rapidly growing popularity of systems approaches in biomedicine makes it increasingly relevant to be able to efficiently compute pairwise statistical associations for \emph{all} available pairs of variables in a dataset. For instance, one could imagine an (online) data explorer where users can interactively mine population cohort data for networks of variables whose pairwise association strengths differ between two user-selected conditions or contexts. Similar network-based data explorers already exist for pre-computed expert-curated networks, e.g., CoVex \cite{Sadegh2020-rx} for exploration of the SARS-CoV-2 virus-host-drug interactome or NeDRex \cite{Sadegh2021-tw} and Drugst.One \cite{Maier2024-ge} for general-purpose exploration of knowledge graphs defined over biomedical entities such as drugs, diseases, and genes. To leverage the data analysis paradigm exemplified in these approaches for the analysis of context-dependent association networks that must be computed on the fly from population cohort data, a statistics tool is required that combines the following two properties:
\begin{itemize}
    \item The tool should be runtime- and memory-efficient enough to enable real-time computation of association networks for large-scale datasets (i.e., computation of all pairwise statistical associations between variables).
    \item The tool should be capable of handling heterogeneous data with high rates of missing values while minimizing information loss (i.e., pairwise data removal instead of complete-case analysis).
\end{itemize} 

As an easily accessible and yet powerful programming language with plenty of available modular extensions, Python has become one of the most popular languages in statistical programming and data mining. Several state-of-the-art Python libraries for statistical analysis exist. While the core Python libraries NumPy \cite{numpy} and pandas \cite{pandas_paper} primarily support basic correlation computations, SciPy \cite{SciPy2020-NMeth} and Pingouin \cite{pingouin} provide a broader spectrum of statistical functions and distributions. However, all of these libraries exhibit specific limitations with respect to the previously described use case. Firstly, pairwise missing data removal or pairwise feature calculations are restricted to a limited subset of statistical tests. Secondly, their application to large-scale datasets within an interactive data explorer is impractical due to the prolonged runtime and the lack of inherent parallelization capabilities. This limitation arises primarily because more efficient popular Python statistics libraries, such as SciPy and pandas, are mainly implemented in Cython, which inherently lacks support for intrinsic parallelization.

For this reason, we developed our Python tool NApy, which addresses all of the above-mentioned issues. We demonstrate the performance of NApy on both simulated data (with varying numbers of features, samples, threads and missing value ratios) and a real-world population cohort. Our results show that NApy uses memory much more efficiently than most available Python competitors and outperforms state-of-the-art competitors in terms of runtime and memory by orders of magnitude.

\section{Related Work}

\begin{table*}
  \caption{Overview of state-of-the-art Python libraries for statistical tests computation including NApy.}
  \label{tab:overview_existing_tools}
  \resizebox{\textwidth}{!}{   
  \begin{tabular}{l p{2.5cm} p{2.5cm} p{2.5cm} p{3cm} p{3cm}}
        \toprule
         & \textbf{NumPy} & \textbf{pandas} & \textbf{SciPy} & \textbf{Pingouin} & \textbf{\nanpy} \\
        \midrule
        Input data & NumPy matrix & pandas DataFrame & NumPy matrix & pandas DataFrame & NumPy matrices \\
        Statistical tests & only $r^2$ & only $r^2$ \& $\rho$ & > 20 & > 20 & 7 \\
        Pairwise input feature calculation & \checkmark & \checkmark & only $\rho$ & only $r^2$ \& $\rho$ & \checkmark \\
        Pairwise missing value removal & $\times$ & \checkmark & only $\rho$ & only $r^2$ \& $\rho$ & \checkmark \\
        P-value calculation & $\times$ & $\times$ & \checkmark & \checkmark & \checkmark \\
        Effect size calculation & $\times$ & $\times$ & limited & extensive & \checkmark \\
        Multiple testing correction & $\times$ & $\times$ & 2 & 5 & 3 \\
        Output format & \makecell[l]{NumPy matrix \\ ($r^2$)} & \makecell[l]{pandas DataFrame \\ ($r^2$ \& $\rho$)} & \makecell[l]{SciPy object \\ (test statistics \& \\ $P$-values)} & \makecell[l]{pandas DataFrame \\ (test statistics, un- \& \\ adjusted $P$-values, \\ effect size, \\ additional results)} & \makecell[l]{dictionary of \\ NumPy arrays \\ (test statistics, un- \& \\ adjusted $P$-values, \\ effect sizes)} \\
        Parallelization & $\times$ & $\times$ & $\times$ & $\times$ & \checkmark \\
        \bottomrule
    \end{tabular}
  }
\end{table*}

A range of computational tools in Python, including NumPy \cite{numpy}, SciPy \cite{SciPy2020-NMeth}, Pingouin \cite{pingouin}, and pandas \cite{pandas_paper, pandas_software_version}, are available for conducting statistical tests, yet each comes with specific limitations concerning our use case (\Cref{tab:overview_existing_tools}). NumPy and pandas, two popular Python libraries for data handling and modeling, are largely limited to correlation analyses and thus only support analyses of continuous data. NumPy is restricted to the calculation of Pearson correlation coefficients and is lacking pairwise deletion of missing data. Conversely, pandas expands correlation options, providing Pearson, Spearman, and Kendall coefficients across features in the data with pairwise deletion of missing values, but does not provide $P$-values. SciPy and Pingouin, on the other hand, offer a broader selection of statistical tests. Expanding upon NumPy’s core features, SciPy offers additional array computation tools, incorporating efficient, low-level language implementations in Fortran, C, and C++ to enhance processing speed. It provides a range of statistical tests for categorical and continuous data, including two-sided $P$-value reporting. Notably, SciPy enables pairwise feature calculations and missing value deletion specifically for Spearman's rank correlation, while SciPy’s other statistical tests lack these capabilities. SciPy further includes multiple testing correction and limited options for effect size calculations. Finally, Pingouin, an open-source statistical package written in Python 3 and based on pandas and NumPy data structures, mostly uses the low-level SciPy functions to provide richer and more exhaustive output. Thus, it includes a wide range of statistical tests with multiple testing correction and effect size calculation directly computed and included in the output pandas DataFrame along with other results, such as the 95\% confidence intervals of the difference in means when calculating a $t$-test. It allows computation of pairwise correlations between columns of a pandas DataFrame with specification of a \texttt{nan\_policy} parameter for pairwise deletion of missing values. However, as in SciPy, these functionalities are only provided for a subset of statistical tests. A common limitation across all four packages in analyzing large-scale datasets with thousands of features and samples is the lack of built-in parallelization.

\par Apart from any considered Python competitors, we also need to mention that to the best of our knowledge, there do not exist any statistics tools in C++ itself suitable for our use case. The Boost library \cite{BoostLibrary} offers functionality for several statistical tests (e.g. $t$-tests and Pearson correlation) but does not offer the computation of statistical tests for all pairwise feature combinations given a data matrix or missing value removal. Similarly, ALGLIB \cite{alglib} offers a higher variability of statistical tests ($t$-tests, $\chi^2$-tests, Mann-Whitney $U$ tests, and correlation coefficients) but does not provide functionality for pairwise feature computations and neglects missing value removal -- furthermore it only supports single-threaded computations in its free version. 

\section{Methodology and Functionality}

\subsection{Overview}\label{subsec:overview}

NApy, available under \url{https://github.com/DyHealthNet/NApy}, consists of seven statistical tests that compute respective statistic values, $P$-values (with support to correct for multiple testing), and effect sizes on given input data matrices in a way that efficiently removes missing values in a pairwise fashion. The implemented tests cover all combinations of continuous, dichotomous, and categorical features, which are the main data types in large biomedical datasets. All tests are implemented in C++, parallelized with OpenMP, and integrated into a Python wrapper by using pybind11 \cite{pybind11}. Additionally, for each test, we offer an analogous implementation based on the Python just-in-time-compiler Numba \cite{numba}. One essential advantage of our implementations is that we perform pairwise missing value removal on-the-fly during statistics computations instead of outsourcing this subsetting procedure as a \enquote{preprocessing} step. 

\subsection{Pairwise Missing Data Removal}\label{subsec:pairwiseRemoval}

One common approach to deal with missing data is to simply restrict to those samples in the data that are fully present (complete-case analysis). However, especially for large datasets and in the presence of high percentages of missingness, this leads to the loss of a significant amount of potentially valuable information \cite{Eekhout2012-im}. On the other hand, standard imputation techniques such as $k$-nearest-neighbor classification, regression imputation, or random forest models provide possibilities to infer missing values \cite{Li2024-zt}. However, the choice of an optimal imputation method can be highly dataset-dependent and downstream analyses can be quite sensitive to the chosen imputation \cite{Shadbahr2023-hb}. An alternative that does not suffer from potential distortion of imputation but on the other hand also does not waste as much information as the complete-case analysis is pairwise missing data removal as implemented in several statistical software libraries (e.g. with the option \texttt{nan\_policy='omit'} in the SciPy library function \texttt{scipy.stats.spearmanr} \cite{SciPy2020-NMeth}). 

\par We briefly want to formalize the concept of pairwise missing data removal as it is implemented in NApy: assume we are given some input data matrix $\data \in \mathbb{R}^{F, S}$ consisting of $F$ features and $S$ samples. Features can be of any type, i.e. either quantitative, dichotomous, or categorical with categories assumed to be label-encoded (i.e. with labels as integers starting from zero). Furthermore, we assume that missing data entries are encoded by some special value, which we will abbreviate by $m$. For a given pair of features $\left(g, h\right)$ with $g, h \in \mathbb{R}^{S}$ (i.e. two rows from $\data$) we want to perform a statistical association analysis. Pairwise missing data removal now means that, for a given feature pair $\left(g, h\right)$, we extract the subset of paired sample entries $\mathcal{I}(g,h)$ where at both positions in $g$ and $h$ the corresponding value is non-missing. More formally, we extract the set of pairs $\mathcal{I}(g, h) := \left\{ \left(g_i, h_i\right) \bigl\vert\ g_i \neq m \land h_i \neq m;\ i=1,\dots,S \right\}$, which is then given as input to the respective statistical test. Note that the actual size of the input set $\mathcal{I}$ depends on the chosen pair of features and on the location of the respective missing data entries. In particular, it may happen that $\mathcal{I}$ is empty.

\subsection{Input}\label{subsec:input}

The input data for all our available statistical tests is stored using two-dimensional objects of the \texttt{numpy.ndarray} class \cite{numpy}. Depending on the respective datatype combination, the test functions expect either one matrix storing only continuous, dichotomous, or categorical feature values, or two input matrices when handling a mix of continuous and categorical or dichotomous feature variables. In either input matrix, a special floating point value indicating missing data can be specified as a parameter by the user. Furthermore, via function parameters the user can specify which input dimension of matrices is supposed to be interpreted as samples and which as features. The number of threads to use in parallel computations of results can also be set via a simple user parameter. Furthermore, the user can specify a list of desired data matrices to be returned such as unadjusted and corrected $P$-values, respective statistics values, and a selection of effect sizes.

\subsection{Output}\label{subsec:output}

For each supported test, the output is represented by a Python dictionary storing data matrices with the user-requested output ($P$-values, adjusted $P$-values, and/or effect sizes). As for the input, we store output matrices as two-dimensional \texttt{numpy.ndarray} objects.  In the case of continuous-continuous and categorical-categorical tests, any returned output matrix $\mathbf{M} \in \mathbb{R}^{F, F}$ for $F$ features in the input data matrix is quadratic and stores at position $\mathbf{M}_{i,j}$ the associated result value for features $i, j \in \left\{1,\dots,F\right\}$. In the heterogeneous case of categorical-continuous input data for $F_{\text{C}}$ categorical and $F_{\text{Q}}$ quantitative input features, such a returned output matrix $\mathbf{M} \in \mathbb{R}^{F_{\text{C}}, F_{\text{Q}}}$ stores at position $\mathbf{M}_{i,j}$ the corresponding result value for categorical feature $i \in \left\{ 1, \dots, F_{\text{C}}\right\}$ and quantitative feature $j \in \left\{1, \dots, F_{\text{Q}}\right\}$. The dichotomous-continuous input and output are defined analogously. 
\par Within our tool, output matrices can either store unadjusted $P$-values, adjusted $P$-values (using either Bonferroni \cite{Dunn1961-hl}, Benjamini-Hochberg \cite{Benjamini1995-fy}, or Benjamini-Yekutieli multiple testing correction \cite{Benjamini2001-qw}), values of the respective test statistic, or suitable effect sizes (Table \ref{table:nanpyOutput}). Note that when applying multiple testing correction for the homogeneous input type case, we do not count \enquote{self-tests} of features on the diagonal as actually performed tests, and therefore set $\mathbf{M}_{i,i} \leftarrow \texttt{numpy.nan}$ for all features $i \in \left\{ 1, \dots, F\right\}$ in any of the multiple testing corrected output matrices $\mathbf{M}$.

\subsection{Implementational Details}

\subsubsection{Parallelization and Missing Data Removal}
All implemented tests have in common that parallelization is employed at the outer level of pairwise feature analysis. That is, the set of all pairs of (either continuous, dichotomous, or categorical) features given some input data matrix (or matrices) is distributed among the number of user-specified threads, which are either generated by OpenMP in the C++ implementation or Numba directly. Hence, in both cases, parallelization runs thread-based on shared memory. No memory needs to be multiplied or copied among processes. In our supplied implementation, for all parametric tests, the Numba implementation is chosen by default, whereas, for nonparametric tests, we use the C++ implementation for efficiency reasons (Section \ref{subsec:resultsSimulated}). However, the user can also specify to use the underlying C++ implementation for parametric tests if desired. For a given pair of features, pairwise missing value removal (as represented by operator $\mathcal{I}$ from Section \ref{subsec:pairwiseRemoval}) is run in combination with the necessary statistical computations. That is, only paired sample entries with present data entries at both positions are kept as input for further calculations. In the following, we briefly sketch individual details for each supplied statistical test.

\subsubsection{Pearson Correlation} 
Two-sided $P$-values for a given pair of features are computed based on the cumulative distribution function (CDF) of the $\beta$-distribution. If the number of remaining sample pairs (i.e. the cardinality of set $\mathcal{I}$) is less than two after pairwise missing data removal, the corresponding $P$-value is not defined and reported as \texttt{numpy.nan}. If the number is at most one, both $r^2$ and $P$-value are not defined and returned as \texttt{numpy.nan}.

\subsubsection{Spearman Correlation} 
First, data are sorted and ranked in parallel for all features in order to avoid costly sorting operations in the later pairwise processing of features. This allows to only account for missing data removal in the pairwise feature processing, which can be implemented in linear time. We account for ties in the data by averaging over tied ranks. Two-sided $P$-values are computed based on the CDF of the $t$-distribution. Similar to Pearson correlation, a remaining sample count of at most two renders $P$-values undefined, and a count of at most one results in both $P$-value and $\rho$ being undefined.

\subsubsection{$\chi^2$-Test} 
$P$-values are calculated based on the survival function of the $\chi^2$-distribution. If any of the previously existing categories becomes empty after the removal of pairwise missing data, the resulting $\chi^2$-value (along with all other potential return values) is undefined and therefore returned as \texttt{numpy.nan}. In case only one category group exists for one of the two categorical features, $P$-value and Cramer's $V$ are undefined and returned as \texttt{numpy.nan}.

\subsubsection{$t$-Test} 
For two-sample $t$-tests, we offer the functionality to let the user choose between applying Student's $t$-test or Welch's $t$-test. The difference is that the latter one does not assume variances of the two populations induced by the dichotomous feature to be equal. Either way, two-sided $P$-values are computed based on evaluating the survival function of the $t$-distribution. If, after pairwise missing value removal, the number of remaining samples in one of the two populations is less than two, or the sum of remaining samples in both categories is less than three, the $t$-statistic is undefined and hence returned as \texttt{numpy.nan} (as well as all other return types). For Student's $t$-test the same holds true in case the pooled standard deviation should evaluate to zero. In Welch's $t$-test, if the sum of the two population variances is zero, Cohen's $D$ is not well-defined and returned as \texttt{numpy.nan}.

\begin{table}
  \centering
  \caption{Overview of NApy's output values for each test.}
  \label{table:nanpyOutput}
  \begin{tabular}{llll}
    \toprule
    \textbf{Test} & \textbf{Statistic} & \textbf{Effect Size(s)} & $P$\textbf{-value(s)} \\ 
    \midrule
    Pearson & & $r^2$ & \multirow{7}{*}{\begin{turn}{90}\begin{tabular}{c} unadjusted \& \\ corrected \\ (Bonf., BH, BY)\end{tabular}\end{turn}} \\ 
    Spearman & & $\rho$ & \\
    $\chi^2$-Test & $\chi^2$ & $\phi$, Cramer’s $V$ & \\
    $t$-Test & $t$ & Cohen’s $D$ & \\ 
    Mann-Whitney $U$ & $U$ & Pearson’s $r$ & \\ 
    ANOVA & $F$ & Partial $\eta^2$ & \\
    Kruskal-Wallis & $H$ & $\eta^2$ & \\ 
    \bottomrule
  \end{tabular}
\end{table}

\subsubsection{Mann-Whitney $U$ Test} 
In order to minimize costly sorting operations during each pairwise analysis of features, each thread initially sorts one continuous variable and subsequently computes test statistics for all dichotomous variables, while accounting for their respective missing values. This way, each continuous variable is sorted only once. We account for ties in the remaining data by averaging over tied ranks and additionally employ the commonly used tie-correction for the computation of $z$-values. Following the Scipy implementation \texttt{scipy.stats.mannwhitneyu}, we offer the possibility to compute $P$-values in an exact or asymptotic way. In the asymptotic mode, $P$-values are computed based on evaluating the survival function of the standard normal distribution. In the exact mode, $P$-values are computed by using a dynamic programming approach presented in \cite{LofflerUnknown-pd}, which is only feasible for smaller input populations. Therefore, we follow the SciPy implementation and offer an \texttt{auto} mode, which automatically chooses between \texttt{asymptotic} and \texttt{exact} mode: if there are no ties for a specific pair of features after missing value removal and any of the two populations has less than eight associated samples, the \texttt{exact} mode is chosen for $P$-value calculation. Otherwise, the \texttt{asymptotic} mode is chosen. If one of the two categories is empty after pairwise missing value removal, the test statistic is not well-defined and we return \texttt{numpy.nan} for all desired return types.

\subsubsection{ANOVA} 
Using ANOVA on mixed type categorical-continuous data, two-sided $P$-values are computed based on the survival function of Fisher's $F$-distribution. As with $t$-tests, if one of the categories after pairwise missing data removal is empty, the test statistic is not well-defined and \texttt{numpy.nan} is returned for all return types. Furthermore, if variances within categories are all zero, we additionally check if all group sums are equal and if not return an $F$-statistic value of \texttt{numpy.infty} and a zero $P$-value. Otherwise, $F$-statistic and $P$-value are undefined and returned as \texttt{numpy.nan}.

\subsubsection{Kruskal-Wallis Test} 
Similar to the Mann-Whitney $U$ test, continuous variables are pre-sorted before pairwise statistical tests in combination with all categorical variables are computed. Ties are treated by averaging over tied ranks and tie correction is used for the computation of the $H$ statistic. In the pairwise feature analysis, after pairwise removal of missing data, we compute two-sided $P$-values based on evaluating the survival function of the $\chi^2$ distribution. In case one of the induced categories is empty after pairwise missing value removal, the test is invalid and \texttt{numpy.nan} is returned for all possible return types. If the number of remaining samples is less than or equal to the number of present categories in the categorical feature, the partial $\eta^2$ value is undefined and returned as \texttt{numpy.nan}.

\subsection{Verification of Correctness}\label{subsec:unit_tests}

We implemented a series of Python-based unit tests to validate the correctness of computed test statistics, $P$-values, and effect sizes. Each test statistic and corresponding $P$-value calculation was benchmarked against results from the SciPy library and equivalent R libraries, such as Hmisc \cite{harrell_jr_hmisc:_2003} and the R stats \cite{R} package using the Python-R-bridge-package rpy2 \cite{rpy2}. For all statistical tests, we tested universally applicable edge cases such as the correctness of the missing value removal functionality, parallel execution, and axis parameter use. Additional tests focused on data-type-specific edge cases, such as single-category data or data without any categories after missing value removal. Finally, effect size calculations and multiple testing corrections were similarly verified against equivalent R implementations with effectsize \cite{ben-shachar_effectsize:_2019}, rstatix \cite{kassambara_rstatix:_2019}, and lsr \cite{navarro_lsr:_2011}. For detailed information on the specific unit tests, please refer to our GitHub repository.

\subsection{Supported System Architectures}

NApy has successfully been installed and run on Linux PCs and servers (Ubuntu 22.04.5, conda 24.9.1) and Mac PCs (macOS Sonoma, conda 24.9.2). For detailed installation instructions on both Linux and Mac, we refer to our GitHub repository.

\section{Benchmarking}

Given our focus on large, mixed-type datasets with a high prevalence of missing values, we assessed runtime and allocated memory usage of NApy, alongside leading state-of-the-art competitors (\Cref{tab:overview_existing_tools}). We used simulated data matrices and the CHRIS study data \cite{Pattaro2015-ok}. All benchmarks were performed on a server within a compute cluster, equipped with two Intel(R) Xeon(R) Silver 4216 CPUs, each operating at 2.10 GHz with 16 physical cores, enabling a total of 64 threads.

\subsection{Setup}

\subsubsection{Simulated Data}

For the first part of our benchmarks, input data matrices of varying sizes were simulated at random. A simulated data matrix $\data \in [0,1]^{F, S}$ with quantitative values is generated by uniformly sampling values from the unit interval $[0,1]$ with $S \in \{ 250, 500, 750, 1000\}$ being the number of samples and $F \in \{ 250, 500, 750, 1000\}$ being the number of features. In the case of categorical data, entries in the matrix represent label-encoded categories, with indices taken from $\{0, ..., c-1\}$, and $c$ being the number of categories for a given feature. We ensured that each feature row contains each possible category at least once. In our benchmark data, we fixed the number of categories for each row to $c = 4$. Given the frequent occurrence of missing values in biomedical data, a predetermined proportion of missing values was introduced at random positions within each feature across both data types. These missing values were represented using a designated floating-point value. For statistical tests that require a single data type, such as Pearson, Spearman, or $\chi^2$-test, a single data matrix was simulated. For tests requiring both quantitative and categorical or dichotomous data, separate data matrices with matching sample and feature dimensions were simulated.

\subsubsection{Real-World Data} 

\begin{figure}[!h]
  \includegraphics[width = \columnwidth]{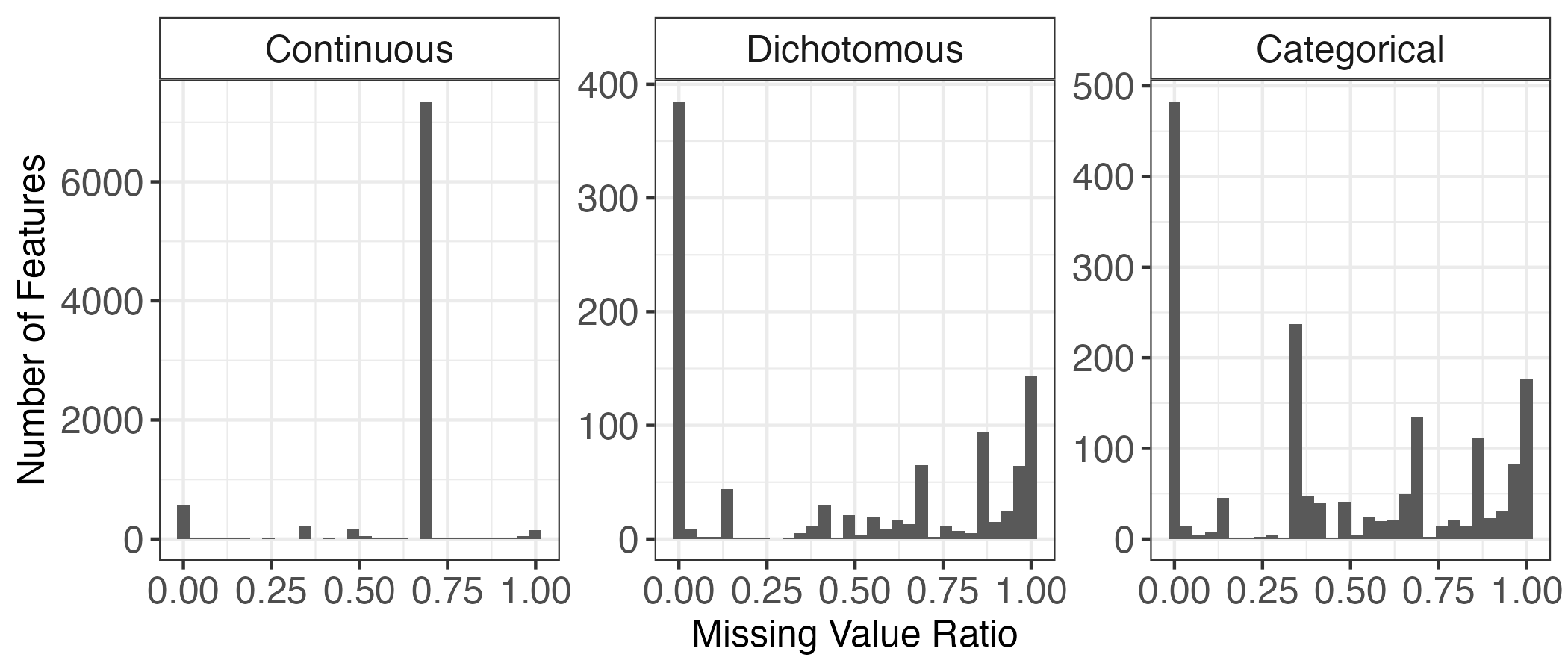}
  \caption{Distribution of missing values in the CHRIS study for continuous, dichotomous, and categorical features.}
  \label{fig:NA_chris_data}
\end{figure}

As real-world data scenario, we used the CHRIS study data, a population-based study cohort of 13,393 adults recruited from 13 municipalities in the alpine Val Venosta/Vinschgau district in the Bolzano-South Tyrol province of northern Italy. Baseline visits were
conducted from 2011 to 2018, collecting socio-demographic, health, lifestyle, and exposure data from questionnaires, interviews, and instrumental examinations, as well
as urine and blood samples for biobanking, DNA extraction, and molecular
characterization (genome and exome sequencing, metabolomics, and proteomics). We use the CHRIS  study data consisting of 10,464 features in total. Specifically, 8803 features are continuous, and 1661 are categorical, among which 1010 features are dichotomous. Additionally, as shown in Figure \ref{fig:NA_chris_data}, a substantial amount of missing data is present in the dataset, primarily resulting from instances where samples were not collected from all patients or questionnaires not answered by all patients.

\subsubsection{Competitor Tools}

Not all state-of-the-art libraries offer the functionalities specific to our application, such as statistical test calculation on all feature pairs and pairwise missing data removal. Therefore, different libraries were selected as competitors, depending on the statistical test. In cases where no suitable tools existed for processing entire data matrices to compute pairwise test statistics and $P$-values, we developed a ``naive'' Python-based baseline. This implementation generates all feature pairs with the help of the Python library itertools and performs calculations iteratively using Numpy data structures. We employed joblib for parallelizing the respective ``naive'' Python baseline implementation, allowing us to compare it to  NApy's runtime performance when run in varying numbers of thread counts.

For Pearson correlation, both Pingouin and pandas provide functions (\texttt{pingouin.rcorr} and \texttt{pandas.DataFrame.corr}) to compute all pairwise correlations across rows or columns of an input data matrix. For Spearman correlation, SciPy supports the computation of all pairwise feature correlations with pairwise deletion of missing values via the function \texttt{scipy.stats.spearmanr}, and we hence included this function in our benchmark. However, unlike NApy, none of the mentioned libraries inherently support parallelization. Consequently, for both Pearson and Spearman correlation, two variants of the Python-based baseline were included, using SciPy's and pandas' correlation coefficient calculations.

For the $\chi^2$-test, the $t$-test, the Mann-Whitney $U$, ANOVA, and the Kruskal-Wallis tests, no direct competitors to NApy are currently available which allow to process all feature pairs via a single function call. Therefore, we only compared NApy against the Python-based baseline implementations of the corresponding functions in the SciPy \texttt{stats} module (\texttt{chi2\_contingency},  \texttt{ttest\_ind}, \texttt{mannwhitneyu}, \texttt{f\_oneway}, \texttt{kruskal}).

It is important to note that the output of the compared competitors and NApy exhibit certain differences. For NApy, we compute and return statistic values and unadjusted $P$-values, as we intended to stay consistent with the output of the competitor SciPy and Pingouin, which we integrated into the benchmarks of all our tests. In contrast, the \texttt{pandas.DataFrame.corr} function only returns a matrix containing correlation values.

\subsubsection{Measuring Runtime and Memory Consumption}

The runtime for each statistical test and library was measured using Python's \texttt{time} module, while the allocated memory of the different implementations was measured using Memray \cite{memray}. Importantly, as stated before, runtime and memory allocation of the competitor pandas was exclusively measured for computing and returning correlations, as pandas does not return $P$-values.

\subsection{Results on Simulated Data}\label{subsec:resultsSimulated}

\subsubsection{Runtime of NApy's C++ and Numba Implementations}

\begin{figure}[!h]
  \includegraphics[width = \columnwidth]{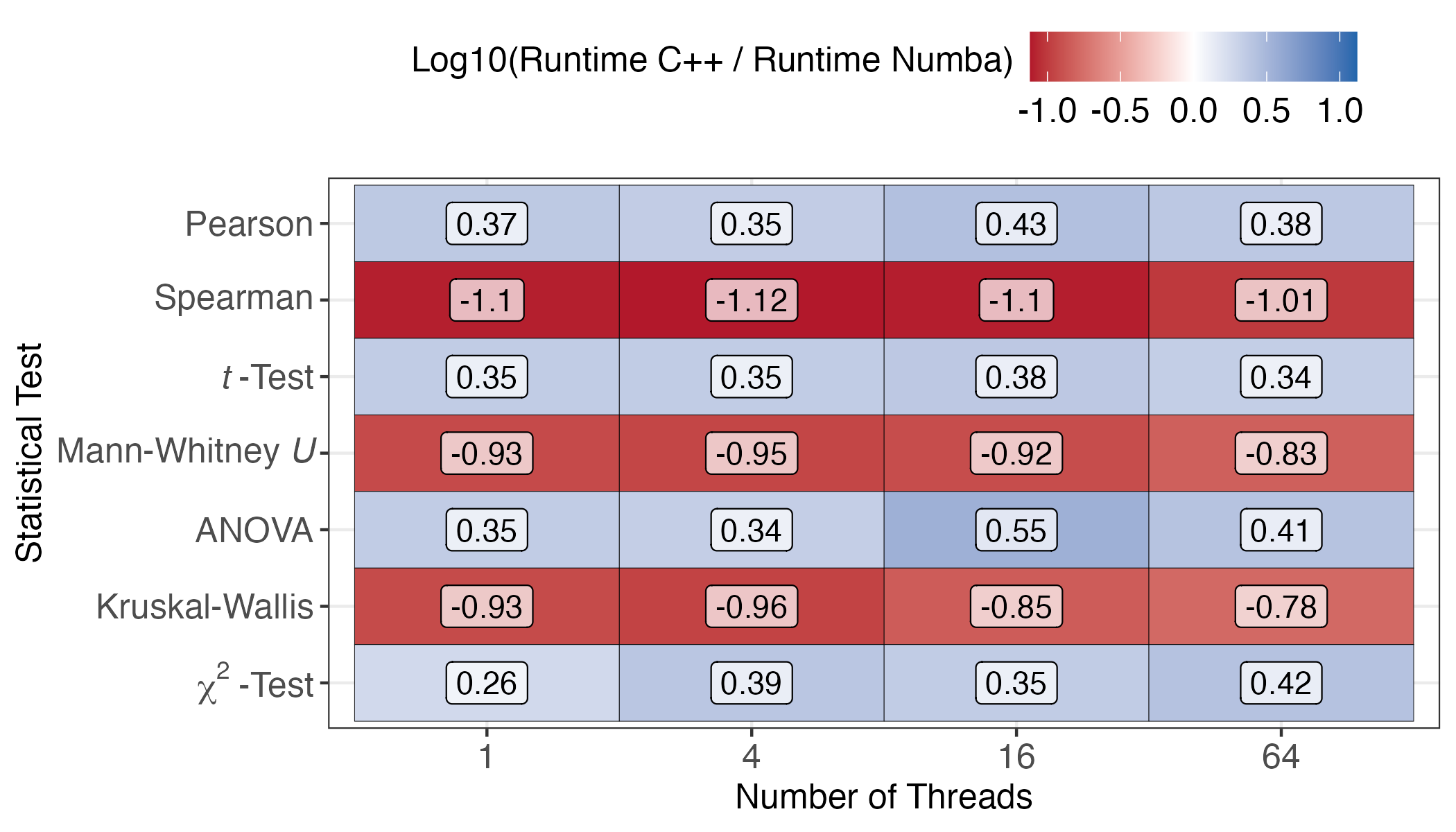}
  \caption{Runtime evaluation of NApy's Numba and C++ implementation of the statistical tests performed using varying numbers of threads on datasets comprising 1000 samples and features with 10\% of missing values per feature. The log10 fold change between the C++ and Numba implementations was computed on the average runtime of three independent runs per statistical test and thread count.}
  \label{fig:numba_C_fold_change}
\end{figure}

For all implemented statistical tests, NApy provides the flexibility to execute computations using either the C++ implementation with OpenMP parallelization or the Numba-parallelized version. This functionality is controlled via the parameter \texttt{use\_numba}, which is integrated into all statistical test functions. We evaluated the runtime of both implementations and compared them by calculating fold changes between the C++ and Numba-based implementations for all statistical tests (Figure \ref{fig:numba_C_fold_change}). The Numba-based implementation was observed to be faster for parametric tests, while C++ code was faster for nonparametric tests. 

\begin{figure*}[!h]
  \includegraphics[width = \textwidth]{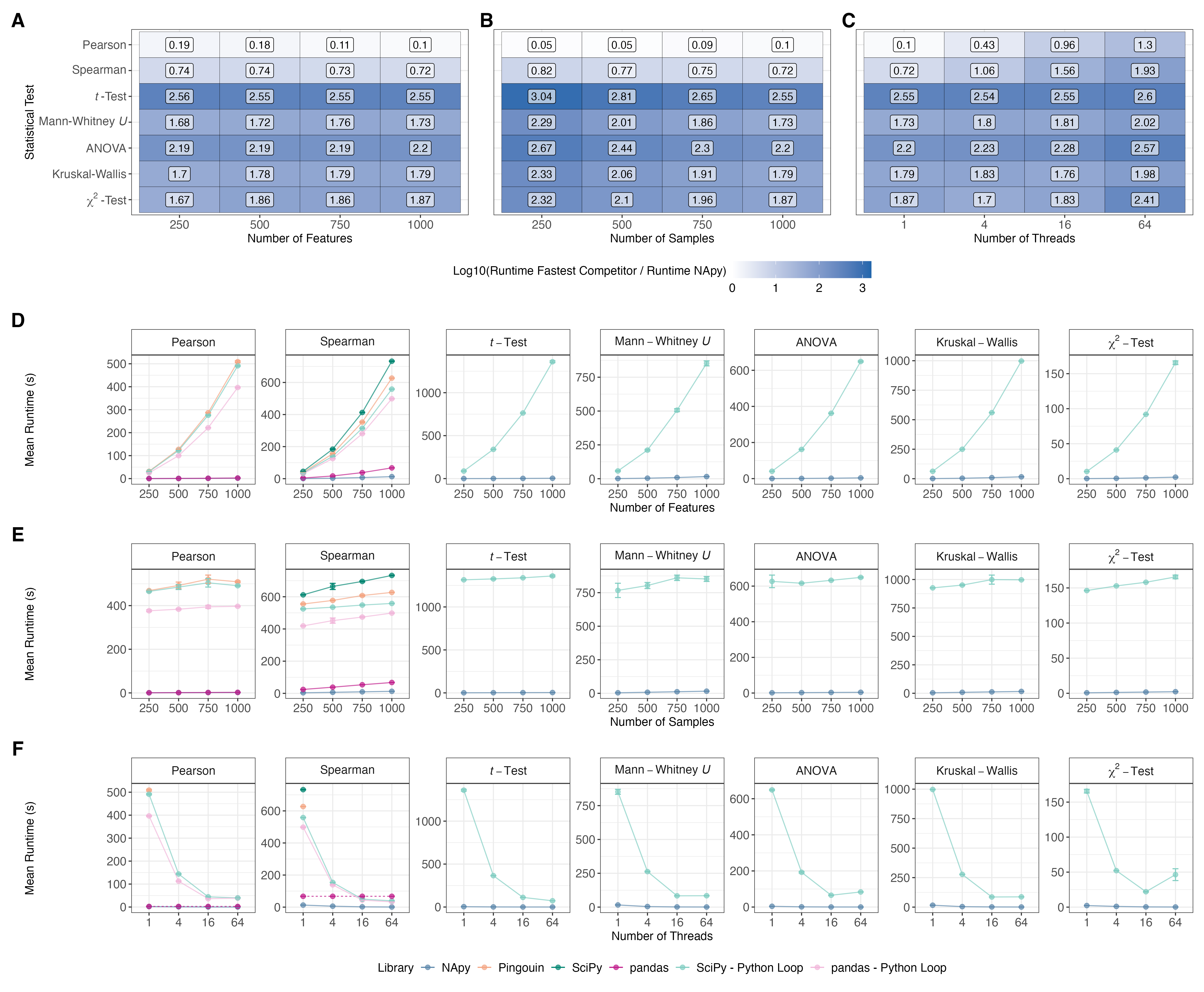}
  \caption{Runtime benchmarks of NApy and applicable competitors across features, sample, and thread counts. (A--C) The log10 fold changes between the runtimes of the fastest competitor (pandas for correlations and Scipy-Python Loop for the remaining tests) and NApy were calculated across the different configurations of features (A), samples (B), and threads (C). (D) The runtime (in seconds) was measured on a single thread for datasets with 1000 samples and varying number of features. (E) Datasets with 1000 features were used to investigate the sample effect on runtime, with computations being performed on a single thread. (F) Datasets of 1000 samples and features were simulated and statistical tests were performed on varying number of threads. As pandas does not support intrinsic parallelization, the dashed line represents its runtime on a single thread. All datasets included 10\% of missing values per feature, and the runtimes were averaged over three independent simulation runs, with standard deviations displayed as error bars.}
  \label{fig:runtime_sample_feature_efffect}
\end{figure*}

Consequently, the default value of this parameter is set to the implementation observed to be faster for each test and from this point forward. References to NApy in all following analyses will correspond to the respective default and hence faster implementation.

\subsubsection{Runtime of NApy Versus Competitors} 

We evaluated NApy's runtimes against available competitors using simulated datasets with varying numbers of features and samples, performing computations across different thread counts (Figure \ref{fig:runtime_sample_feature_efffect}). For all tests and all numbers of features, samples, and threads, NApy is consistently faster than all tested competitors (Figure \ref{fig:runtime_sample_feature_efffect}A--C). In the vast majority of the tested configurations, the speedup is massive; in the most extreme case, NApy is more than 1000 times faster than its fastest competitor (Figure \ref{fig:runtime_sample_feature_efffect}B).

The only scenarios where comparable runtimes can be achieved with competitor tools are correlation calculations on a single thread (first two rows in Figure \ref{fig:runtime_sample_feature_efffect}A--B). Here, the fastest available competitor is pandas which, especially for computing Pearson correlation coefficients, is only slightly slower than NApy. Moreover, when NApy is allowed to run in several threads, it clearly outperforms all competitors also for the correlation coefficient computations (first two rows in Figure \ref{fig:runtime_sample_feature_efffect}C), highlighting its improved scaling behavior in comparison to existing tools.

Unsurprisingly, we observed that the number of features has a greater impact on runtime compared to the number of samples (Figure \ref{fig:runtime_sample_feature_efffect}D--E), aligning with expectations since an increase in features leads to a quadratic increase in the number of pairwise statistical tests. Similarly, thread count significantly impacts runtime, with higher thread counts reducing runtime as anticipated due to well-scaling parallelization (Figure \ref{fig:runtime_sample_feature_efffect}F).

\subsubsection{Runtime Under Varying Missing Value Ratios}

\begin{figure}[!h]
  \includegraphics[width = \columnwidth]{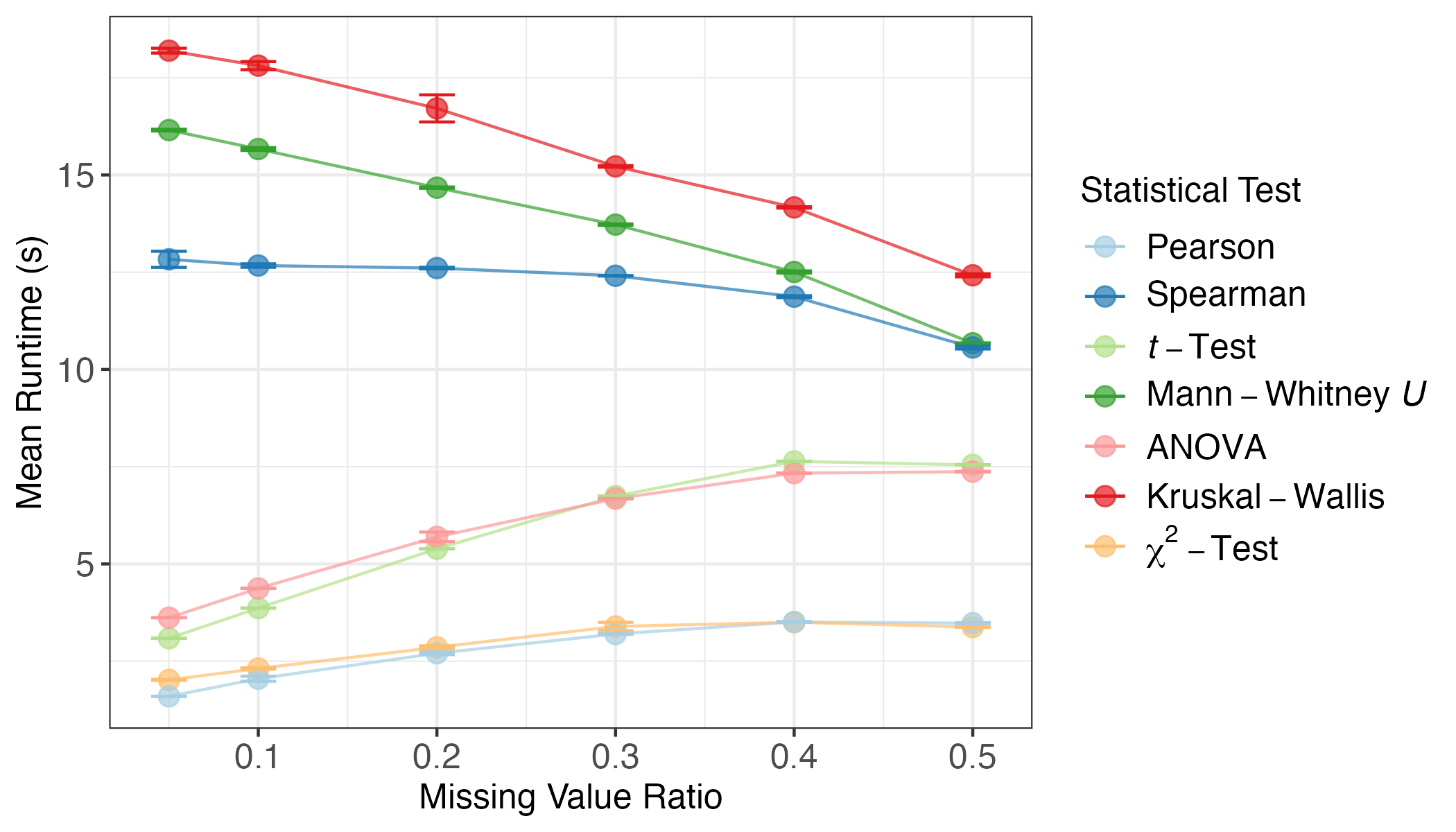}
  \caption{Benchmark analysis of the impact of missing values per feature on the runtime of statistical tests in NApy. Simulated datasets with fixed sizes of 1000 features and 1000 samples were analyzed to evaluate the effect of varying levels of missing values per feature on runtime. All computations were performed using a single thread, and runtime measurements were averaged over three runs, with standard deviations represented as error bars.}
  \label{fig:na_ratio_pearson_spearman}
\end{figure}

Another relevant dataset characteristic influencing the number of available data entries is the proportion of missing data. We assessed its impact by measuring runtime of NApy across varying rates of missing values per feature, using fixed-sized datasets with 1000 samples and features. The analysis revealed that, for parametric tests, runtime increased with the proportion of missing values in the dataset, whereas, for nonparametric tests, an inverse relationship was observed (Figure \ref{fig:na_ratio_pearson_spearman}). This behavior can mainly be attributed to the fact that the sorting and ranking routines of the nonparametric tests rely on more elaborate data structures, which become smaller and cheaper to maintain as the number of available data points decreases.

\begin{table*}[!h]
  \centering
  \caption{Comparison of allocated memory usage between NApy and state-of-the-art competitors on a fixed dataset size of 1000 features and samples with 10\% missingness per feature. Memory consumption (in GB) for all statistical tests and libraries was averaged over three runs.}
  \label{tab:memoryBenchmark}
  \begin{tabular}{lllllllll}
    \toprule
    \textbf{Library} & \textbf{Threads} &\textbf{Pearson} & \textbf{Spearman} & \textbf{$t$-Test} & \textbf{Mann-Whitney $U$} & \textbf{ANOVA} & \textbf{Kruskal-Wallis} & \textbf{$\chi^2$-Test} \\
    \midrule
    \multirow{4}{*}{NApy} & 1 & 0.015 & \underline{\textbf{3.825}} & \underline{\textbf{0.263}} & \underline{\textbf{0.079}} & \underline{\textbf{0.289}} & \underline{\textbf{0.132}} & \underline{\textbf{0.297}} \\
      & 4 & 0.015 & 3.825 & 0.263 & 0.079 & 0.289 & 0.132 & 0.297 \\
       & 16 & 0.015 & 3.825 & 0.263 & 0.079 & 0.289 & 0.132 & 0.297 \\
        & 64 & 0.015 & 3.825 & 0.263 & 0.079 & 0.289 & 0.132 & 0.297 \\
    Pingouin & 1 & 124.248 & 250.447 & & & & & \\
    SciPy& 1 & & 190.244  & & & & & \\
    pandas& 1 & \underline{\textbf{0.009}}  & 37.671 & & & & & \\
    SciPy - Python Loop& 1 & 129.025 & 226.225 & 244.166  & 221.865 & 131.943 & 235.246 & 61.965 \\
    pandas - Python Loop& 1 & 37.935 & 99.91 & & & & & \\
    \bottomrule
  \end{tabular}
\end{table*}

\subsubsection{Allocated Memory of NApy Versus Competitors} 
We benchmarked allocated memory usage of NApy in comparison competitors for a fixed dataset size of $1000$ features and $1000$ samples with a fixed ratio of missing values per feature of $0.1$ and varying number of threads used by NApy (Table \ref{tab:memoryBenchmark}). Our measurements show  that, except for the computation of Pearson correlation coefficients (which is anyways rather cheap in terms of absolute memory consumption), NApy is clearly more memory-efficient than all other tested tools. Due to the shared-memory-parallelization based on OpenMP and Numba, the memory consumption of NApy remains unchanged even when more than one thread is used. This is a clear advantage in contrast to Python-parallelized competitors, as, due to the Global Interpreter Lock, efficient parallelization in Python is mainly based on multiprocessing which leads to the multiplication of memory under the usage of more than one process.

\subsection{Results on Real-World Data}

To demonstrate the practical applicability of NApy on a real-world dataset, all statistical tests were performed on the CHRIS study data both without any parallelization (single thread) and with heavy parallelization (64 threads). For correlations, pandas was used as comparison as it demonstrated to be the fastest compared to other competitors on the simulated data (Figure 4). The results show that, for computing Pearson correlation using a single thread, pandas slightly outperformed NApy in terms of runtime efficiency. However, due to pandas' lack of parallelization capability, NApy achieved significantly improved runtime performance when utilizing 64 threads, with fold changes of 8.32 ($\log_{10}(8.32) = 0.92$) and 24.98 ($\log_{10}(24.98) = 1.4$) observed for Pearson and Spearman correlations, respectively. For the remaining statistical tests, the runtime of SciPy-Python Loop was compared to that of NApy, and we observed that, on a single thread as well as on 64 threads, NApy consistently achieved better runtimes than SciPy. Specifically, with NApy, runtime improvement fold changes ranged from 2 to 14 on a single thread, and even reached maximum values of over 400 using 64 threads, which emphasizes the drastic speedup that NApy allows on such large datasets. 

\begin{figure}[!h]
  \includegraphics[width = \columnwidth]{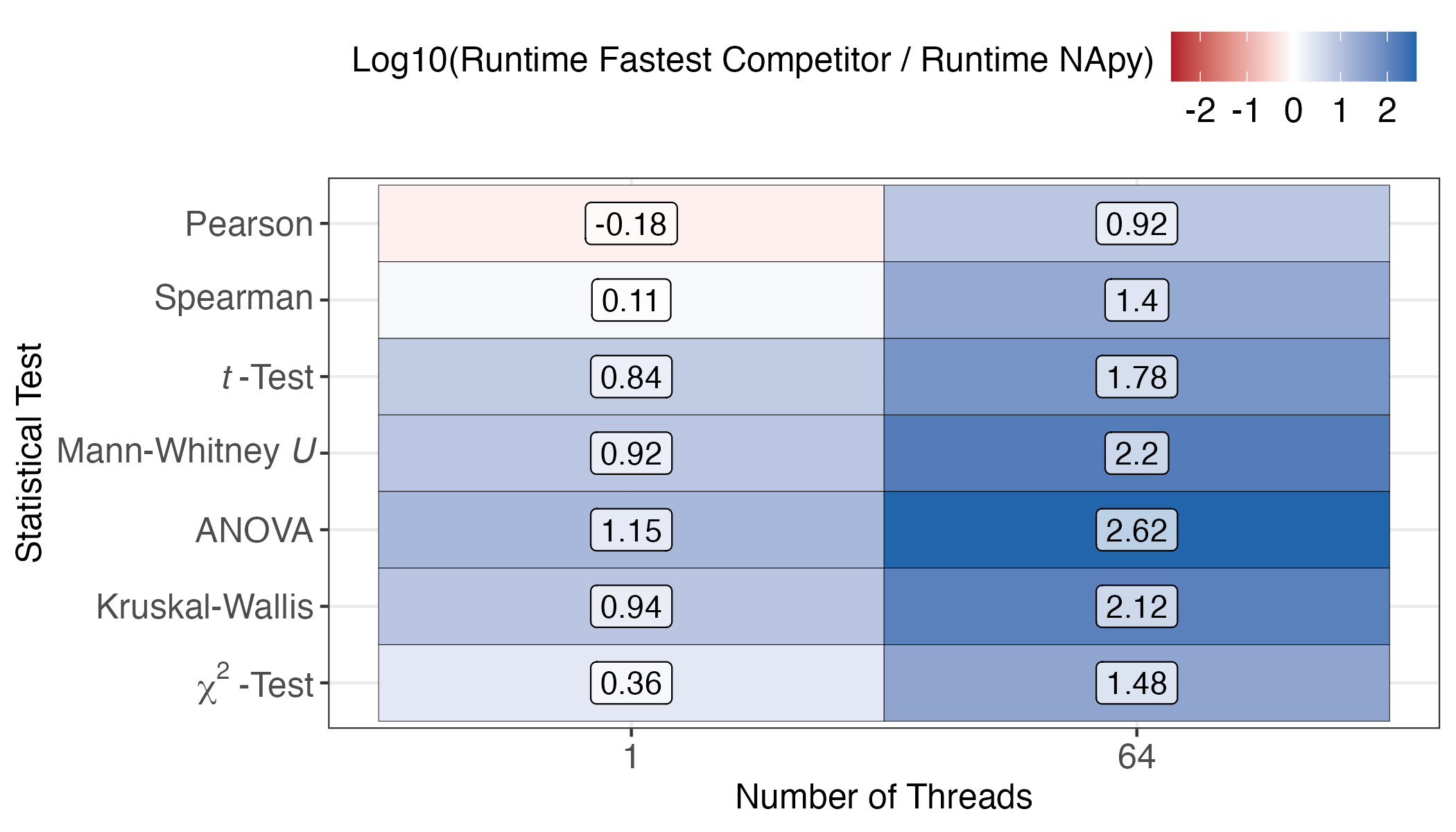}
  \caption{Runtime evaluation of NApy and its fastest competitor on the CHRIS study data. The comparison includes pandas for correlation tests and Scipy-Python Loop for other statistical tests. The log10 fold change was derived from the average runtime of three independent runs per statistical test and thread count.}
  \label{fig:chris_fold_changes}
\end{figure}

\section{Limitations}\label{sec:limitations}
The first and probably most obvious limitation of NApy is the number of statistical tests that are currently implemented, especially in contrast to well-established Python statistic libraries such as SciPy, and Pingouin. However, NApy is currently focused on the most commonly employed statistical tests such that it can already be used in practice for the analysis of mixed-type data in biomedical data analysis and beyond. Especially, the application of statistical tests accounting for covariates would be beneficial for analyses involving population-based cohort data. Secondly, one has to note that in direct comparison to e.g. the Pingouin library, NApy currently does not offer the possibility to return as much (meta-)information on statistical results. Yet, we believe that for users working on such large-scale datasets with the focus on efficiency and speed, our current output in the form of statistics values, effect sizes, and (multiple-testing corrected) $P$-values offers a rich source of information well-suited for data analysis focused on especially large datasets. Lastly, we need to mention that for the special case of computing Pearson correlation in the absence of missing data, in the sequential case it can definitely be useful to resort to the NumPy implementation, which largely benefits from internal vectorization. However, as soon as missing data comes into play and more elaborate statistical tests need to be applied, NumPy can no longer offer suitable functionality.

\section{Conclusion}

In this paper, we have presented our Python tool NApy, which is capable of performing statistical tests on large heterogeneous input data in the presence of missing data efficiently in parallel. Existing Python statistic tools currently lack the functionality to compute statistical tests on large input data while accounting for pairwise missing data removal in parallel. This becomes an issue as soon as data analyses on heterogeneous large-scale datasets need to be run in a time-critical context. 
\par On both simulated and real-world population-based cohort data, we show that NApy consistently outperforms Python competitors and competing Python baseline implementations in terms of runtime efficiency and memory allocation. An exception arised in the computation of Pearson correlation using single thread, where pandas showed reduced memory allocation on simulated data and faster execution on cohort data. It is important to note, however, that pandas only computes correlation coefficients, whereas NApy also performs $P$-value calculations, which were not factored into the runtime and memory allocation comparisons.
\par We are convinced that NApy will enable the development of new real-time statistical analyses on large datasets stemming from experimental measurements that are currently still limited to analyses on precomputed statistical evaluations. 

\section{Data and Code Availability}
The source code of NApy can be accessed at \url{https://github.com/DyHealthNet/NApy}, while benchmark results and implementations are available at \url{https://github.com/DyHealthNet/NApy_benchmark}. Data and samples of the CHRIS study can be requested for clearly defined research via the CHRIS Portal (\url{https://chrisportal.eurac.edu/}).


\begin{acks}

The work was supported by the Deutsche Forschungsgemeinschaft (DFG, German Research Foundation) -- 516188180.

The CHRIS study was funded by the Autonomous Province of Bolzano-South Tyrol - Department of Innovation, Research, University and Museums and supported by the European Regional Development Fund (FESR1157). CHRIS study investigators thank all study participants, the Healthcare System of the Autonomous Province of Bolzano-South Tyrol, and all Eurac Research staff involved in the study. Bioresource Impact Factor Code: BRIF6107.

The Ethics Committee of the Healthcare System of the Autonomous Province of Bolzano-South Tyrol approved the CHRIS baseline protocol on 19 April 2011 (21-2011). The study conforms to the Declaration of Helsinki, and with national and institutional legal and ethical requirements.

\end{acks}

\bibliographystyle{ACM-Reference-Format}
\bibliography{references}

\end{document}